\documentclass[aps,prb,reprint,preprintnumbers,showpacs,superscriptaddress,amsmath,amssymb]{revtex4-1}

\usepackage{graphicx}% Include figure files
\usepackage{dcolumn}% Align table columns on decimal point
\usepackage{bm}% bold math

\begin{document}

%\preprint{}
%\preprint{\textit{Optical Properties of Microcavity with fluctuations}}

\title{Evaluation of two-photon polarization density matrix of polarization-entangled photon-pairs generated through biexciton resonant hyper-parametric scattering}
% Force line breaks with \\

\author{Y. Yamamoto}
\affiliation{Department of Physical Science, Graduate School of Science, Osaka Prefecture University, 
1-1 Gakuen-cho, Naka-ku, Sakai 599-8531, JAPAN}
\author{G. Oohata}
\affiliation{Department of Physical Science, Graduate School of Science, Osaka Prefecture University, 
1-1 Gakuen-cho, Naka-ku, Sakai 599-8531, JAPAN}
\author{K. Mizoguchi}
\affiliation{Department of Physical Science, Graduate School of Science, Osaka Prefecture University, 
1-1 Gakuen-cho, Naka-ku, Sakai 599-8531, JAPAN}

\date{\today}% It is always \today, today,

\begin{abstract}
We have investigated the excitation-power dependence of polarization-entangled photon-pairs generated from a CuCl single crystal using biexciton resonant hyper-parametric scattering.
The measured two-photon polarization density matrix which corresponds to the two-photon polarization state changes with increasing the excitation power.
The evaluation of the tangle and the linear entropy obtained from the measured density matrix indicates that the two-photon polarization density matrix can be expressed as the mixed state of an ideal state and a totally mixed state, and the mixture ratio of the totally mixed state increases as the excitation density increases.
The variations of the density matrix and the mixture ratio with the excitation power originate from the generation of the multiple photon pairs.
\end{abstract}

%\keywords{Multilayers, Polaritons, Excitons and related phenomena}
%

\maketitle

\section{\label{sec:Intro} Introduction}

The study for the generation of quantum entangled photon pair is an important subject to progress quantum optics and quantum information \cite{Bouwmeester1997,Kok2007}.
The spontaneous parametric down conversion (SPDC) is generally utilized as the generation method of quantum entangled photon pairs \cite{Kwiat1995}.
In recent years, the generation methods by the utilization of the third order nonlinear optical phenomena have been reported \cite{Wang2001,Sharping2006}.
These methods have the features as the enhancement of the effective generation area of photon pairs \cite{Li2009} and the generation of the photon pairs in the wavelength range of the telecommunication bands \cite{Dong2014}.
In the third-order nonlinear phenomena, the generation method by using biexciton resonant hyper-parametric scattering (RHPS) which is induced by the resonance of the two-photon transition to a discrete energy state show the high generation efficiency of entangled photon pairs \cite{Edamatsu2004,Oohata2007}.
Furthermore, in RHPS, because the characteristics of the excited quantum state in materials reflect the generated entangled photon pair, the properties of entangled photon pairs are hardly affected by the conditions of excitation light.
We consider that the generation method using RHPS is suitable to apply to the non-classical light source with high brightness such as entanglement laser \cite{Simon2003}.
However, there are few reports on  generation method using RHPS, and the quantum states under the condition of the generation of multiple photon pairs are not discussed both theoretically and experimentally. Therefore, the investigation of the excitation-power dependence of quantum states of photon pairs is an important subject to clarify the quantum states under the generation of the multiple photon pairs.
We have been studied the generation method of entangled photon pairs by using RHPS, and reported that the generation rates of photon pairs and the generation efficiency of photon pair per unit excitation power have been estimated by two-photon time-correlation measurements \cite{Yamamoto2016}.
However, the quantum states of the photon pairs under various observation conditions have been not discussed and the excitation-power dependence of the quantum state has been not clarified.
We believe that it is important to clarify the quantum state of photon pairs under the high power excitation.
%Then, in this study, we have investigated the excitation-power dependence of the quantum state of photon pairs by performing two-polarization projection measurements for polarized-entangled photon-pair generated by RHPS.

In this paper, we report on the excitation-power dependence of the quantum states of entangled photon pairs emitted from CuCl crystals by RHPS.
We have measured the polarization projection for photon pairs generated by RHPS at various excitation powers.
Two-photon polarization density matrices have been reconstructed from the experimental results.
%Two-photon polarization density matrices have been reconstructed from the experimental results obtained by the polarization projection measurement.
The fidelities of the ideal state with the measured density matrices have been evaluated.
Furthermore, we have estimated the linear entropy and tangle from the measured density matrix of entangled photon pairs and clarified the relationship between the linear entropy and the tangle at various excitation powers.
%Furthermore, we have first estimated the linear entropy and tangle from the measured density matrix of entangled photon pairs and clarified the relationship between the linear entropy and the tangle at various excitation powers.
The obtained relationship between the linear entropy and the tangle is consistent with that in the Werner state, which can be explained by taking account of multiple photon pairs due to the increase in the generation rate of photon pairs.

%\section{Experiment}
%\begin{figure}[tbp]
%\centering
%\includegraphics[width= 4.5 cm]{image/figure1_v1.eps}
%\caption{Schematic diagram of experimental setup for the resonant hyper-parametric scattering (RHPS) method. ND: neutral density filter, PA: polarization analyzer, $\mathrm{M_A}$ and $\mathrm{M_B}$: monochromators, PMT: Photomultiplier tube. }
%\label{fig:setup}
%\end{figure}

\section{\label{sec:Expeimental} Experiment and analysis}

A CuCl single crystal sample with a size of approximately $5 \times 5 \times 0.1\ \mathrm{mm^3}$ grown from vapor phase was used to generate the entangled photon pairs by RHPS.
The temperature of the sample was maintained at 10 K in a cryostat.
The pump pulses were the second harmonic light of a mode-locked Ti:Sapphire pulse laser at a repetition rate of 80 MHz.
For the excitation condition, the center wavelength and the spectral width of the pump pulses were selected to $\lambda$=390 nm and $\Delta\lambda$= 0.2 nm, respectively, by using a 4f optical system composed of two lenses, two gratings, and a slit.
The center wavelength of the pump pulses corresponds to the two-photon resonance of the biexciton state in CuCl.
The pump pulses passing through a neutral density filter were focused on the sample, where the spot size of the pump pulses on the sample was approximately 100 $\mu \mathrm{m} \phi$.
The emitted light in the directions of about 45 and -45 degrees from the excitation optical axis were taken into the multi-mode fibers and guided to monochromators.
The polarized components were selected through a polarization analyzer composed of quarter wave plates, half wave plates, and polarizing plates at the front of the fibers.
To satisfy a phase matching condition of RHPS, the detected wavelength of the RHPS light emitted from CuCl was selected by monochromators, and the RHPS light was detected using single-photon counting photomultipliers.
The photon counting signals were input to a time correlator, and the coincidence counts were measured.
Moreover, the polarization projection measurements were carried out at various excitation powers.
According to the method reported by Kwiat {\it et al}. \cite{James2001}, the quantum state tomography for two-photon polarization density matrix was obtained by measuring 16 kinds of polarized projection.

\section{\label{sec:Results} Results and Discussion}

\begin{figure}[tbp]
\centering
\includegraphics[width=8 cm]{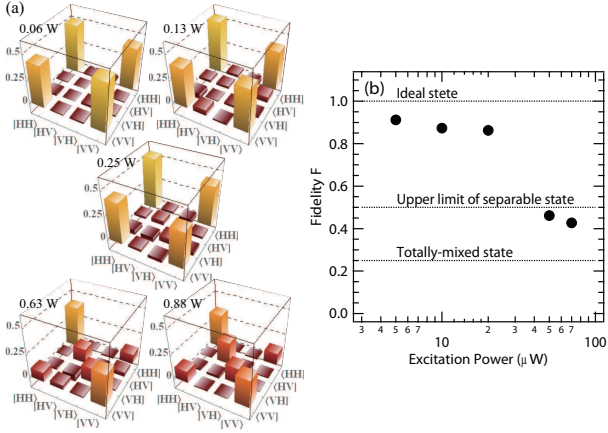}
\caption{(a) The real parts of the measured density matrices for the two-photon polarization states generated from RHPS method at various excitation densities. (b) Excitation-power dependence of Fidelity $F$. Dashed lines indicate the expected value of the fidelity for ideal state, an upper limit of separable state, and totally mixed state.}
\label{fig:DM}
\end{figure}
The ideal state for polarization entangled photon pairs generated by RHPS is represented as \cite{Edamatsu2004}
\begin{equation}
\left| {{\psi _i}} \right\rangle  = \frac{1}{{\sqrt 2 }}\left( {\left| {HH} \right\rangle  + \left| {VV} \right\rangle } \right),
\label{eq:idealstate}
\end{equation}
where H and V indicate horizontal and vertical linear polarization, respectively. The density matrix $\rho_{i}$ for the ideal state is given by 
\begin{equation}
{\rho _i} = \left| {{\psi _i}} \right\rangle \left\langle {{\psi _i}} \right| = \begin{pmatrix}
1/2 		&0		&0		&1/2		\\
0		&0 		&0		&0		\\
0		&0		&0 		&0		\\
1/2		&0		&0		&1/2
\end{pmatrix}.
\label{eq:idealDM}
\end{equation}
Moreover, the fidelity $F$ of the ideal state with the measured density matrix $\rho$ obtained from the quantum state tomography is defined as
 \begin{equation}
F = \left\langle {{\psi _i}|\rho |{\psi _i}} \right\rangle .
\label{eq:DefFidelity}
\end{equation}
To clarify the excitation-power dependence of the fidelity for the entangled photon pairs generated by RHPS, the two-photon polarization density matrices have been investigated at various excitation powers.
The two-photon polarization density matrices at various excitation powers, which are reconstructed from the results of the polarization projection measurements, were shown in Fig. \ref{fig:DM} (a). It is found that when the excitation power is the lowest, the measured density matrix is similar to the ideal state.
The fidelity of the ideal state with the measured density matrix at the lowest excitation power is 0.91 which is almost consistent with the value reported by Oohata {\it et al}. \cite{Oohata2007}.
Moreover, the measured density matrices are changed with increasing the excitation intensity.
As the excitation power is increased, the non-diagonal components $\left| {HH} \right\rangle \left\langle {VV} \right|$and$\left| {VV} \right\rangle \left\langle {HH} \right|$ of the measured density matrices are decreased and the diagonal components$\left| {HV} \right\rangle \left\langle {HV} \right|$ and$\left| {VH} \right\rangle \left\langle {VH} \right|$ are increased.
In order to quantitatively evaluate the deviation from the ideal state with the increase in the excitation power, the fidelity $F$ was calculated at various excitation powers (Fig. \ref{fig:DM} (b)).
It is clear that the fidelity is monotonously decreased with increasing the excitation power.
The fidelity in the range larger than about 50$\ \mu$W is smaller than the classical upper limit of the fidelity, which indicates that the quantum entanglement between photons in the photon pair emitted from CuCl completely disappears.

	The tangle $T$ and the linear entropy $S_L$ as the physical quantities \cite{Wootters1998, Bose2000} are obtained to reveal the origin of the excitation-power dependence of the density matrix. The tangle $T$ represents the degree of entanglement.%which is defined as 
%\begin{equation}
%T = {\left[ {\max \left( {{\lambda _1} - {\lambda _2} - {\lambda _3} - {\lambda _4},0} \right)} %\right]^2}.
%\label{eq:DefTangle}
%\end{equation}
%Here, $\lambda_{i} (i=1,2,3,4)$ is the number obtained by arranging the square roots of the eigenvalues of the matrix, $R = \rho ({\sigma ^y} \otimes {\sigma ^y}){\rho ^*}({\sigma ^y} \otimes {\sigma ^y})$, in descending order. $\rho$ and $\sigma^{y}$ are the measured density matrix and Pauli matrix, respectively.
%The operator   means the direct product of matrices.
The tangle $T$ takes values from 0 to 1, and indicates that the degree of entanglement is higher as the tangle becomes close to 1.
The linear entropy $S_L$ represents the mixedness of the density matrix.
%which is defined as
%\begin{equation}
%{S_L} = \frac{4}{3}\left( {1 - Tr\left[ {{\rho ^2}} \right]} \right).
%\label{eq:DefSL}
%\end{equation}
%where $Tr[ ]$ indicates taking the trace of the matrix.
The linear entropy $S_L$ takes values from 0 to 1.
$S_L$ takes 0 in the pure state, and shows a value close to 1 as many states are mixed.
\begin{figure}[tbp]
\centering
\includegraphics[width=6 cm]{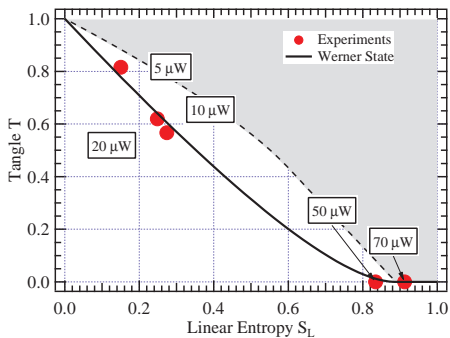}
\caption{ Relationship between Linear entropy $S_L$ and Tangle $T$. Red circles indicate the experimental values of $S_L$ and $T$ obtained from the measured density matrices at various excitation powers. Solid line indicates the Werner states and gray area corresponds to the states that are not physically realized.}
\label{fig:phase}
\end{figure}
The relationship between the tangle $T$ and the linear entropy $S_L$ obtained at various excitation powers is plotted in Fig. \ref{fig:phase}.
The dashed line indicates the maximum entangled state \cite{Wootters1998}, and the gray area shows the physically impossible states.
As the excitation power is increased, the measured density matrix changes from the state with the high tangle and low entropy to one with the low tangle and high entropy.
From the obtained relationship between $T$ and $S_L$, we predict that the density matrix under the high excitation power constructs the mixed state consisting of the ideal state and the totally mixed state, which is called Werner state \cite{Munro2001}.
Using an ideal density matrix $\rho_{i}$ and a totally mixed state $\rho_{TM}=I/4$, where $I$ is the unit matrix, the Werner state $\rho_{W}$ is defined as 
\begin{equation}
\rho_W = (1 - g) \rho_i + g \rho_{TM}.
\label{eq:WernerState}
\end{equation}
Here, $g$ is the mixture ratio of the ideal state and the totally mixed state, which takes values from 0 to 1.
Moreover, it is known that the Werner state monotonically changes from the state with low entropy and high tangle to one with high entropy and low tangle states as the parameter $g$ changes from 0 to 1.
The solid line in Fig. \ref{fig:phase} shows the relationship between $T$ and $S_L$ in the Werner state obtained by Eq. (\ref{eq:WernerState}).
The experimental results almost agree with the solid line.
This agreement means that the measured density matrix obeys the Werner state, which indicates that the parameter $g$ is increased with increasing the excitation power; namely, the proportion of the totally mixed state increases with the increase of the excitation power. 

\begin{figure}[tbp]
\centering
\includegraphics[width=6 cm]{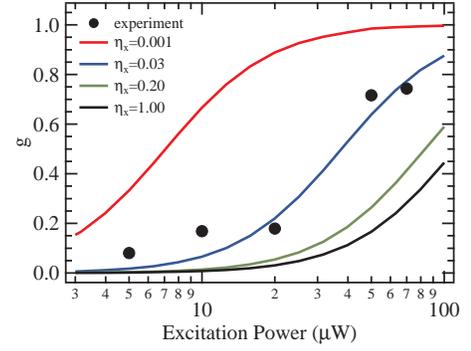}
\caption{Excitation-power dependence of $g$. Black circles indicate the obtained value of $g$ from the measured density matrices at various excitation powers. Solid lines indicate calculated excitation-power dependence of $g$ at various values of the simultaneous detection efficiency $\eta_{x}$ (0.001, 0.03, 0.20 and 1.00).}
\label{fig:gvalue}
\end{figure}

To understand the effect of the excitation power on the measured density matrix, it is necessary to clarify the origin of the totally mixed state.
We have obtained the parameters $g$ at various excitation powers by fitting the measured density matrix to the Werner state.
The excitation-power dependence of the parameter $g$ is shown in Fig. \ref{fig:gvalue}.
The experimental results marked with black circles indicate that the parameter $g$ is monotonously increased with the increase of the excitation power, which indicates that the proportion of the totally mixed state depends on excitation power. 

When it is assumed that the origin of the totally mixed state is due to the photoluminescence by bound excitons in CuCl and the scattering light of the pump pulses \cite{Oohata2007}, the detection rate of the totally mixed state is proportional to the coincidence count rate of two photons which depends on the square of the excitation power, because the photoluminescence by bound excitons and the scattering of the pump pulses belong to a linear optical phenomenon.
On the other hand, in RHPS which is related to a third-order nonlinear parametric process, the detection rate of photon pairs is proportional to the square of the excitation power.
Therefore, the ratio of the two-photon coincidence count rate by RHPS to that by the photoluminescence and the scattering takes constant values with the excitation power.
This property for the ratio of the two-photon coincidence count rate has been experimentally confirmed in the previous report \cite{James2001}.
Moreover, this result means that the parameter $g$ caused by the photoluminescence and scattering is independent on excitation power, which indicates that the obtained excitation-power dependence of the parameter $g$ will not originate from the photoluminescence by bound excitons and the scattering of the pump pulses.
Therefore, we consider that the change of the measured density matrix and the excitation-power dependence of the parameter $g$ will be caused by multiple photon pairs.

As the excitation power increases, the generation rate of photon pairs by RHPS increases, and thus the probability of multiple photon pairs simultaneously generated also increases.
When the multiple photon pairs are generated, the measured density matrix is represented by the base of number states corresponding to the number of generated photon pairs.
In this case, to completely clarify the measured density matrix, we need at least twice as the number of detectors as the number of photon pairs \cite{Simon2003}.
In the present study, because the coincidence count rates have been observed by using two detectors, the information up to only two photons can be obtained. When three or more photons are generated, it is considered that the polarized projection which is statistically obtained differs from the correlation of only two photons.
Takesue {\it et al}. have theoretically calculated the effect of the generation of multiple photon pairs on polarization correlation measurements with two detectors \cite{Takesue2010}.
When multiple photon pairs are generated independently, the coincidence probabilities $R_{HH}$, $R_{HV}$, and $R_{HR}$ for the respective polarization projections HH, VV, and HR are expressed by 
\begin{eqnarray}
\label{eq:TakesueHH}
R_{HH} &=& \sum\limits_{x = 0}^{\infty} f \left( x \right) P_p \left( {x,\mu } \right) \simeq {\alpha ^2}\left( {\frac{\mu }{2} + \frac{\mu ^2}{4}} \right),\\[-4pt]
\label{eq:TakesueHV}
{R_{HV}} &=& \sum\limits_{x = 0}^{\infty} {g\left( x \right){P_p}\left( {x,\mu } \right)} \simeq {\alpha ^2}\frac{{{\mu ^2}}}{4},\\[-4pt]
\label{eq:TakesueHR}
{R_{HR}} &=& \sum\limits_{x = 0}^{\infty} {h\left( x \right){P_p}\left( {x,\mu } \right)} \simeq {\alpha ^2}\left( {\frac{\mu }{4} + \frac{{{\mu ^2}}}{4}} \right),
\end{eqnarray}
where $x$, $P_{p}$, $\alpha$ and $\mu$ are the number of photon pairs generated simultaneously, the Poisson distribution function, the detection efficiency of photons, the average photon pair number, respectively.
$f(x)$,$g(x)$, and $h(x)$ expressed as follows are the all coincidence probabilities of detection of photons of the polarization projections HH, HV and HR, respectively, when $x$ pairs are generated.
%$f(x)$,$g(x)$, and $h(x)$ expressed as follows are the coincidence probabilities of the polarization projections HH, HV and HR, respectively, when $x$ pairs are generated.
\begin{alignat}{2}
\label{eq:TakesueF}
f(x) =& \sum\limits_{y = 0}^{x} & \frac{1}{2^x} \frac{x!}{y!(x-y)!} & \left\{1-(1-\alpha)^{x-y} \right\}^2, \\[-3pt]
%\end{eqnarray}
%\vspace{-18pt}
%\begin{eqnarray}
\label{eq:TakesueG}
g(x) =& \sum\limits_{y = 0}^{x} & \frac{1}{2^x} \frac{x!}{y!(x-y)!} & \left\{1-(1-\alpha)^{x-y} \right\} \nonumber \\[-3pt]
\ & & & \times \left\{1-(1-\alpha)^{y} \right\},
\end{alignat}
%\end{eqnarray}
%\vspace{-18pt}
\begin{eqnarray}
\label{eq:TakesueH}
h(x) &=& \left(\sum\limits_{j = 0}^{x}  \frac{1}{2^x} \frac{x!}{j!(x-j)!}  \left\{\{1-(1-\alpha)\right\}^{j} \right)\nonumber \\[-3pt]
\ \!\!\!& & \times \left(\sum\limits_{l = 0}^{x}  \frac{1}{2^x} \frac{x!}{l!(x-l)!}  \left\{\{1-(1-\alpha)\right\}^{l} \right).
\end{eqnarray}
The two-photon polarization density matrix obtained from these polarized projections is expressed as \cite{James2001}
\begin{eqnarray}
\rho_{p}(\mu) &\simeq& \frac{1}{4+4\mu}
\begin{pmatrix}
2+\mu 	&0		&0		&2		\\
0		&\mu 	&0		&0		\\
0		&0		&\mu 	&0		\\
2		&0		&0		&2+\mu
\end{pmatrix} \nonumber \\[-4pt]
\ &=& \frac{1}{{1 + \mu }}{\rho _i} + \frac{\mu }{{1 + \mu }}{\rho _{TM}}.
\label{eq:TakesueDM}
\end{eqnarray}
This equation indicates that the density matrix can be expressed by the mixed state of the ideal state and the totally mixed state, namely, the Werner state.
Moreover, the parameter $g$ is expressed by $g = \mu/(1+\mu)$, which is monotonously increased with increasing $\mu$.
On the other hand, since the simultaneous detection efficiency $\eta_{x}$ affects the visibility of the time-correlation signal, it is expected that $\eta_{x}$ will have a significant influence on the calculation of the density matrix \cite{James2001}.
Then, the detection probabilities for the polarized projections, $R'_{HH}$, $R'_{HV}$, and $R'_{HR}$, are represented as
\begin{eqnarray}
\label{eq:HH}
{R'_{HH}} &=& \sum\limits_{x = 0}^{{N_{\max }}} {f'\left( {x,{\eta _x}} \right){P_p}\left( {x,\mu } \right)} ,\\[-4pt]
\label{eq:HV}
{R'_{HV}} &=& \sum\limits_{x = 0}^{{N_{\max }}} {g'\left( {x,{\eta _x}} \right){P_p}\left( {x,\mu } \right)} ,\\ [-4pt]
\label{eq:HR}
{R'_{HR}} &=& \sum\limits_{x = 0}^{{N_{\max }}} {h'\left( {x,{\eta _x}} \right){P_p}\left( {x,\mu } \right)} , 
\end{eqnarray}
by taking account of $\eta_{x}$ based on Eq. (\ref{eq:TakesueHH}) - (\ref{eq:TakesueHR}).
Here, to carry out an nummerical calculation to these polarizxation projections, we put $N_{max}$ as an upper limit of the number of the simultaneous generated photon-pairs.
%Here, $N_{max}$, and $P_{p}$ indicate the upper limit of the number of the simultaneous generated photon-pairs and the Poisson distribution function, respectively.
%x indicates the number of photon pairs generated simultaneously.
$f'(x, \eta_{x})$, $g'(x, \eta_{x})$ and $h'(x, \eta_{x})$ expressed as follows are the coincidence probabilities of the polarization projections HH, HV and HR, respectively, when $x$ pairs are observed with the simultaneous detection efficiency $\eta_{x}$. 
\begin{alignat}{1}
  f'(x,{\eta _x}) =& \mathop \sum \limits_{k = 0}^x \mathop \sum \limits_{m = 0}^{x - k} X(x,k,m,{\eta _x})f(x,k,m)  \\[-4pt]
  g'(x,{\eta _x}) =& \mathop \sum \limits_{k = 0}^x \mathop \sum \limits_{m = 0}^{x - k} X(x,k,m,{\eta _x})g(x,k,m)  \\[-4pt]
  h'(x,{\eta _x}) =& \mathop \sum \limits_{k = 0}^x \mathop \sum \limits_{m = 0}^{x - k} X(x,k,m,{\eta _x})h(x,k,m) 
\label{eq:TermEtaX}
\end{alignat}
where,
\begin{equation}
X(x,k,m,{\eta _x})  = \eta _x^k{\left( {\frac{{1 - {\eta _x}}}{2}} \right)^{x - k}}\frac{{x!}}{{k!(x - k - m)!m!}}, \hfill
\end{equation}
\vspace{-18pt}
\begin{alignat}{1}
f(x,k,m) =& \mathop \sum \limits_{y = 0}^k \sum \limits_{z = 0}^{x - k - m}  \sum \limits_{w = 0}^m \nonumber \\
\ & \left( \frac{1}{2} \right)^x \frac{k!(x - k - m)!m!}{(k - y)!y!(x - k - m - z)!z!(m - w)!w!}  \nonumber \\
\ & \times \left\{ 1 - (1 - \alpha )^{y + z} \right\} \left\{ 1 - (1 - \alpha )^{y + w} \right\}, \\ %\label{eq:TermSumF}
%\end{eqnarray}
%\vspace{-18pt}
%\begin{eqnarray}
%\label{eq:TermSumG}
g(x,k,m) =& \sum \limits_{y = 0}^k  \sum \limits_{z = 0}^{x - k - m} \sum \limits_{w = 0}^m  \nonumber \\
\ & \left( \frac{1}{2} \right)^x \frac{k!(x - k - m)!m!}{(k - y)!y!(x - k - m - z)!z!(m - w)!w!}  \nonumber \\
\ & \times \left\{ 1 - (1 - \alpha )^{y + z} \right\} \left\{ 1 - (1 - \alpha )^{k - y + m - w} \right\}, \\
%\end{eqnarray}
%\vspace{-18pt}
%\begin{eqnarray}
%\label{eq:TermSumH}
h(x,k,m) =& \left( \sum \limits_{y = 0}^{x - m} \!\! \frac{1}{2^{x - m}} \! \frac{(x - m)!}{(x - m - y)!y!} \!\! \left\{ \!1 - (1 - \alpha )^{x - m - y} \! \right\} \!\!\! \right) \nonumber \\
\ & \times \!\! \left( \sum \limits_{z = 0}^{k + m} \!\! \frac{1}{2^{k + m}} \! \frac{(k + m)!}{(k + m - z)!z!} \!\! \left\{ \!1 - (1 - \alpha )^{k + m - z} \! \right\} \!\!\! \right).
\end{alignat}
Using Eq. (\ref{eq:HH}) - (\ref{eq:HR}), the polarization projections are calculated at various values of $\eta_{x}$, and thus the two-photon polarization density matrices are obtained by performing the quantum state tomography.
In this calculation, we adopted $N_{max}=15$.
The solid lines in Fig. \ref{fig:gvalue} indicate the calculated parameter $g$ plotted as function of excitation power at various values of $\eta_{x}$.
The excitation-power dependence of $g$ drastically changes with the value of $\eta_{x}$.
At $\eta_{x} = 0.03$, the calculated excitation-power dependence of $g$ is in almost agreement with the experimental results.
%However, the calculated excitation-power dependence slightly deviates from the experimental results in the region of the low excitation power.
%This small deviation will be induced by the background photons due to the photoluminescence by bound excitons and the scattering of the pump pulses, as mentioned above.
%This small deviation will be induced by the background photons due to the photoluminescence by bound excitons and the scattering of the pump pulses, which indicates that the parameter $g$ is independent on excitation power, as mentioned above.
From these results, we demonstrate that the excitation-power dependence of the measured density matrix originates from the generation of multiple photon pairs.
Moreover, the fact that the mixture ratio of the ideal state and the totally mixed state $g$ depends on excitation power can be explained by taking account of the generation of multiple photon pairs and the simultaneous detection efficiency $\eta_{x}$. 

\section{\label{sec:Conclusion} Conclusion}
We have investigated the polarization projections of the photon pairs generated by RHPS at various excitation powers.
The two-photon polarization density matrices, which are reconstructed from the results of the polarization projection measurements, drastically change with the excitation powers.
The fidelities of the ideal state with the measured density matrices are monotonously decreasing with increasing the excitation power.
The relationship between the tangle and linear entropy obtained at various excitation powers indicates that the measured density matrix under the high excitation powers reflects the mixed state consisting of the ideal state and the totally mixed state.
Moreover, the proportion of the totally mixed state increases with the increase in the excitation power, which originates from the generation of multiple photon pairs.

This work was supported by Research Foundation for Opto-Science and Technology and Support Center for Advanced Telecommunications Technology Research.

%\newpage %Just because of unusual number of tables stacked at end
%\begin{references}

\bibliography{forOL}

\begin{thebibliography}{16}%
\makeatletter
\providecommand \@ifxundefined [1]{%
 \@ifx{#1\undefined}
}%
\providecommand \@ifnum [1]{%
 \ifnum #1\expandafter \@firstoftwo
 \else \expandafter \@secondoftwo
 \fi
}%
\providecommand \@ifx [1]{%
 \ifx #1\expandafter \@firstoftwo
 \else \expandafter \@secondoftwo
 \fi
}%
\providecommand \natexlab [1]{#1}%
\providecommand \enquote  [1]{``#1''}%
\providecommand \bibnamefont  [1]{#1}%
\providecommand \bibfnamefont [1]{#1}%
\providecommand \citenamefont [1]{#1}%
\providecommand \href@noop [0]{\@secondoftwo}%
\providecommand \href [0]{\begingroup \@sanitize@url \@href}%
\providecommand \@href[1]{\@@startlink{#1}\@@href}%
\providecommand \@@href[1]{\endgroup#1\@@endlink}%
\providecommand \@sanitize@url [0]{\catcode `\\12\catcode `\$12\catcode
  `\&12\catcode `\#12\catcode `\^12\catcode `\_12\catcode `\%12\relax}%
\providecommand \@@startlink[1]{}%
\providecommand \@@endlink[0]{}%
\providecommand \url  [0]{\begingroup\@sanitize@url \@url }%
\providecommand \@url [1]{\endgroup\@href {#1}{\urlprefix }}%
\providecommand \urlprefix  [0]{URL }%
\providecommand \Eprint [0]{\href }%
\providecommand \doibase [0]{http://dx.doi.org/}%
\providecommand \selectlanguage [0]{\@gobble}%
\providecommand \bibinfo  [0]{\@secondoftwo}%
\providecommand \bibfield  [0]{\@secondoftwo}%
\providecommand \translation [1]{[#1]}%
\providecommand \BibitemOpen [0]{}%
\providecommand \bibitemStop [0]{}%
\providecommand \bibitemNoStop [0]{.\EOS\space}%
\providecommand \EOS [0]{\spacefactor3000\relax}%
\providecommand \BibitemShut  [1]{\csname bibitem#1\endcsname}%
\let\auto@bib@innerbib\@empty
%</preamble>
\bibitem [{\citenamefont {Bouwmeester}\ \emph {et~al.}(1997)\citenamefont
  {Bouwmeester}, \citenamefont {Pan}, \citenamefont {Mattle}, \citenamefont
  {Eibl}, \citenamefont {Weinfurter},\ and\ \citenamefont
  {Zeilinger}}]{Bouwmeester1997}%
  \BibitemOpen
  \bibfield  {author} {\bibinfo {author} {\bibfnamefont {D.}~\bibnamefont
  {Bouwmeester}}, \bibinfo {author} {\bibfnamefont {J.-w.}\ \bibnamefont
  {Pan}}, \bibinfo {author} {\bibfnamefont {K.}~\bibnamefont {Mattle}},
  \bibinfo {author} {\bibfnamefont {M.}~\bibnamefont {Eibl}}, \bibinfo {author}
  {\bibfnamefont {H.}~\bibnamefont {Weinfurter}}, \ and\ \bibinfo {author}
  {\bibfnamefont {A.}~\bibnamefont {Zeilinger}},\ }\href {\doibase
  10.1038/37539} {\bibfield  {journal} {\bibinfo  {journal} {Nature}\ }\textbf
  {\bibinfo {volume} {390}},\ \bibinfo {pages} {575} (\bibinfo {year}
  {1997})}\BibitemShut {NoStop}%
\bibitem [{\citenamefont {Kok}\ \emph {et~al.}(2007)\citenamefont {Kok},
  \citenamefont {Nemoto}, \citenamefont {Ralph}, \citenamefont {Dowling},\ and\
  \citenamefont {Milburn}}]{Kok2007}%
  \BibitemOpen
  \bibfield  {author} {\bibinfo {author} {\bibfnamefont {P.}~\bibnamefont
  {Kok}}, \bibinfo {author} {\bibfnamefont {K.}~\bibnamefont {Nemoto}},
  \bibinfo {author} {\bibfnamefont {T.~C.}\ \bibnamefont {Ralph}}, \bibinfo
  {author} {\bibfnamefont {J.~P.}\ \bibnamefont {Dowling}}, \ and\ \bibinfo
  {author} {\bibfnamefont {G.~J.}\ \bibnamefont {Milburn}},\ }\href {\doibase
  10.1103/RevModPhys.79.135} {\bibfield  {journal} {\bibinfo  {journal} {Rev.
  Mod. Phys.}\ }\textbf {\bibinfo {volume} {79}},\ \bibinfo {pages} {135}
  (\bibinfo {year} {2007})}\BibitemShut {NoStop}%
\bibitem [{\citenamefont {Kwiat}\ \emph {et~al.}(1995)\citenamefont {Kwiat},
  \citenamefont {Mattle}, \citenamefont {Weinfurter}, \citenamefont
  {Zeilinger}, \citenamefont {Sergienko},\ and\ \citenamefont
  {Shih}}]{Kwiat1995}%
  \BibitemOpen
  \bibfield  {author} {\bibinfo {author} {\bibfnamefont {P.}~\bibnamefont
  {Kwiat}}, \bibinfo {author} {\bibfnamefont {K.}~\bibnamefont {Mattle}},
  \bibinfo {author} {\bibfnamefont {H.}~\bibnamefont {Weinfurter}}, \bibinfo
  {author} {\bibfnamefont {A.}~\bibnamefont {Zeilinger}}, \bibinfo {author}
  {\bibfnamefont {A.}~\bibnamefont {Sergienko}}, \ and\ \bibinfo {author}
  {\bibfnamefont {Y.}~\bibnamefont {Shih}},\ }\href {\doibase
  10.1103/PhysRevLett.75.4337} {\bibfield  {journal} {\bibinfo  {journal}
  {Phys. Rev. Lett.}\ }\textbf {\bibinfo {volume} {75}},\ \bibinfo {pages}
  {4337} (\bibinfo {year} {1995})}\BibitemShut {NoStop}%
\bibitem [{\citenamefont {Wang}\ \emph {et~al.}(2001)\citenamefont {Wang},
  \citenamefont {Hong},\ and\ \citenamefont {Friberg}}]{Wang2001}%
  \BibitemOpen
  \bibfield  {author} {\bibinfo {author} {\bibfnamefont {L.~J.}\ \bibnamefont
  {Wang}}, \bibinfo {author} {\bibfnamefont {C.~K.}\ \bibnamefont {Hong}}, \
  and\ \bibinfo {author} {\bibfnamefont {S.~R.}\ \bibnamefont {Friberg}},\
  }\href {\doibase 10.1088/1464-4266/3/5/311} {\bibfield  {journal} {\bibinfo
  {journal} {J. Opt. B Quantum Semiclassical Opt.}\ }\textbf {\bibinfo {volume}
  {3}},\ \bibinfo {pages} {346} (\bibinfo {year} {2001})}\BibitemShut {NoStop}%
\bibitem [{\citenamefont {Sharping}\ \emph {et~al.}(2006)\citenamefont
  {Sharping}, \citenamefont {Lee}, \citenamefont {Foster}, \citenamefont
  {Turner}, \citenamefont {Schmidt}, \citenamefont {Lipson}, \citenamefont
  {Gaeta},\ and\ \citenamefont {Kumar}}]{Sharping2006}%
  \BibitemOpen
  \bibfield  {author} {\bibinfo {author} {\bibfnamefont {J.~E.}\ \bibnamefont
  {Sharping}}, \bibinfo {author} {\bibfnamefont {K.~F.}\ \bibnamefont {Lee}},
  \bibinfo {author} {\bibfnamefont {M.~A.}\ \bibnamefont {Foster}}, \bibinfo
  {author} {\bibfnamefont {A.~C.}\ \bibnamefont {Turner}}, \bibinfo {author}
  {\bibfnamefont {B.~S.}\ \bibnamefont {Schmidt}}, \bibinfo {author}
  {\bibfnamefont {M.}~\bibnamefont {Lipson}}, \bibinfo {author} {\bibfnamefont
  {A.~L.}\ \bibnamefont {Gaeta}}, \ and\ \bibinfo {author} {\bibfnamefont
  {P.}~\bibnamefont {Kumar}},\ }\href {\doibase 10.1364/OE.14.012388}
  {\bibfield  {journal} {\bibinfo  {journal} {Opt. Express}\ }\textbf {\bibinfo
  {volume} {14}},\ \bibinfo {pages} {12388} (\bibinfo {year}
  {2006})}\BibitemShut {NoStop}%
\bibitem [{\citenamefont {Li}\ \emph {et~al.}(2009)\citenamefont {Li},
  \citenamefont {Yang}, \citenamefont {Ma}, \citenamefont {Cui}, \citenamefont
  {Ou},\ and\ \citenamefont {Yu}}]{Li2009}%
  \BibitemOpen
  \bibfield  {author} {\bibinfo {author} {\bibfnamefont {X.}~\bibnamefont
  {Li}}, \bibinfo {author} {\bibfnamefont {L.}~\bibnamefont {Yang}}, \bibinfo
  {author} {\bibfnamefont {X.}~\bibnamefont {Ma}}, \bibinfo {author}
  {\bibfnamefont {L.}~\bibnamefont {Cui}}, \bibinfo {author} {\bibfnamefont
  {Z.~Y.}\ \bibnamefont {Ou}}, \ and\ \bibinfo {author} {\bibfnamefont
  {D.}~\bibnamefont {Yu}},\ }\href {\doibase 10.1103/PhysRevA.79.033817}
  {\bibfield  {journal} {\bibinfo  {journal} {Phys. Rev. A}\ }\textbf {\bibinfo
  {volume} {79}},\ \bibinfo {pages} {033817} (\bibinfo {year}
  {2009})}\BibitemShut {NoStop}%
\bibitem [{\citenamefont {Dong}\ \emph {et~al.}(2014)\citenamefont {Dong},
  \citenamefont {Zhou}, \citenamefont {Zhang}, \citenamefont {He},
  \citenamefont {Zhang}, \citenamefont {You}, \citenamefont {Huang},\ and\
  \citenamefont {Peng}}]{Dong2014}%
  \BibitemOpen
  \bibfield  {author} {\bibinfo {author} {\bibfnamefont {S.}~\bibnamefont
  {Dong}}, \bibinfo {author} {\bibfnamefont {Q.}~\bibnamefont {Zhou}}, \bibinfo
  {author} {\bibfnamefont {W.}~\bibnamefont {Zhang}}, \bibinfo {author}
  {\bibfnamefont {Y.}~\bibnamefont {He}}, \bibinfo {author} {\bibfnamefont
  {W.}~\bibnamefont {Zhang}}, \bibinfo {author} {\bibfnamefont
  {L.}~\bibnamefont {You}}, \bibinfo {author} {\bibfnamefont {Y.}~\bibnamefont
  {Huang}}, \ and\ \bibinfo {author} {\bibfnamefont {J.}~\bibnamefont {Peng}},\
  }\href {\doibase 10.1364/OE.22.000359} {\bibfield  {journal} {\bibinfo
  {journal} {Opt. Express}\ }\textbf {\bibinfo {volume} {22}},\ \bibinfo
  {pages} {359} (\bibinfo {year} {2014})}\BibitemShut {NoStop}%
\bibitem [{\citenamefont {Edamatsu}\ \emph {et~al.}(2004)\citenamefont
  {Edamatsu}, \citenamefont {Oohata}, \citenamefont {Shimizu},\ and\
  \citenamefont {Itoh}}]{Edamatsu2004}%
  \BibitemOpen
  \bibfield  {author} {\bibinfo {author} {\bibfnamefont {K.}~\bibnamefont
  {Edamatsu}}, \bibinfo {author} {\bibfnamefont {G.}~\bibnamefont {Oohata}},
  \bibinfo {author} {\bibfnamefont {R.}~\bibnamefont {Shimizu}}, \ and\
  \bibinfo {author} {\bibfnamefont {T.}~\bibnamefont {Itoh}},\ }\href {\doibase
  10.1038/nature02838} {\bibfield  {journal} {\bibinfo  {journal} {Nature}\
  }\textbf {\bibinfo {volume} {431}},\ \bibinfo {pages} {167} (\bibinfo {year}
  {2004})}\BibitemShut {NoStop}%
\bibitem [{\citenamefont {Oohata}\ \emph {et~al.}(2007)\citenamefont {Oohata},
  \citenamefont {Shimizu},\ and\ \citenamefont {Edamatsu}}]{Oohata2007}%
  \BibitemOpen
  \bibfield  {author} {\bibinfo {author} {\bibfnamefont {G.}~\bibnamefont
  {Oohata}}, \bibinfo {author} {\bibfnamefont {R.}~\bibnamefont {Shimizu}}, \
  and\ \bibinfo {author} {\bibfnamefont {K.}~\bibnamefont {Edamatsu}},\ }\href
  {\doibase 10.1103/PhysRevLett.98.140503} {\bibfield  {journal} {\bibinfo
  {journal} {Phys. Rev. Lett.}\ }\textbf {\bibinfo {volume} {98}},\ \bibinfo
  {pages} {140503} (\bibinfo {year} {2007})},\ \Eprint
  {http://arxiv.org/abs/0607139v2} {arXiv:0607139v2 [arXiv:quant-ph]}
  \BibitemShut {NoStop}%
\bibitem [{\citenamefont {Simon}\ and\ \citenamefont
  {Bouwmeester}(2003)}]{Simon2003}%
  \BibitemOpen
  \bibfield  {author} {\bibinfo {author} {\bibfnamefont {C.}~\bibnamefont
  {Simon}}\ and\ \bibinfo {author} {\bibfnamefont {D.}~\bibnamefont
  {Bouwmeester}},\ }\href {\doibase 10.1103/PhysRevLett.91.053601} {\bibfield
  {journal} {\bibinfo  {journal} {Phys. Rev. Lett.}\ }\textbf {\bibinfo
  {volume} {91}},\ \bibinfo {pages} {1} (\bibinfo {year} {2003})}\BibitemShut
  {NoStop}%
\bibitem [{\citenamefont {Yamamoto}\ \emph {et~al.}(2016)\citenamefont
  {Yamamoto}, \citenamefont {Oohata},\ and\ \citenamefont
  {Mizoguchi}}]{Yamamoto2016}%
  \BibitemOpen
  \bibfield  {author} {\bibinfo {author} {\bibfnamefont {Y.}~\bibnamefont
  {Yamamoto}}, \bibinfo {author} {\bibfnamefont {G.}~\bibnamefont {Oohata}}, \
  and\ \bibinfo {author} {\bibfnamefont {K.}~\bibnamefont {Mizoguchi}},\ }\href
  {\doibase 10.1364/OE.24.006034} {\bibfield  {journal} {\bibinfo  {journal}
  {Opt. Express}\ }\textbf {\bibinfo {volume} {24}},\ \bibinfo {pages} {6034}
  (\bibinfo {year} {2016})}\BibitemShut {NoStop}%
\bibitem [{\citenamefont {James}\ \emph {et~al.}(2001)\citenamefont {James},
  \citenamefont {Kwiat}, \citenamefont {Munro},\ and\ \citenamefont
  {White}}]{James2001}%
  \BibitemOpen
  \bibfield  {author} {\bibinfo {author} {\bibfnamefont {D.~F.~V.}\
  \bibnamefont {James}}, \bibinfo {author} {\bibfnamefont {P.~G.}\ \bibnamefont
  {Kwiat}}, \bibinfo {author} {\bibfnamefont {W.~J.}\ \bibnamefont {Munro}}, \
  and\ \bibinfo {author} {\bibfnamefont {A.~G.}\ \bibnamefont {White}},\ }\href
  {\doibase 10.1103/PhysRevA.64.052312} {\bibfield  {journal} {\bibinfo
  {journal} {Phys. Rev. A}\ }\textbf {\bibinfo {volume} {64}},\ \bibinfo
  {pages} {052312} (\bibinfo {year} {2001})}\BibitemShut {NoStop}%
\bibitem [{\citenamefont {Wootters}(1998)}]{Wootters1998}%
  \BibitemOpen
  \bibfield  {author} {\bibinfo {author} {\bibfnamefont {W.~K.}\ \bibnamefont
  {Wootters}},\ }\href {\doibase 10.1103/PhysRevLett.80.2245} {\bibfield
  {journal} {\bibinfo  {journal} {Phys. Rev. Lett.}\ }\textbf {\bibinfo
  {volume} {80}},\ \bibinfo {pages} {2245} (\bibinfo {year}
  {1998})}\BibitemShut {NoStop}%
\bibitem [{\citenamefont {Bose}\ and\ \citenamefont {Vedral}(2000)}]{Bose2000}%
  \BibitemOpen
  \bibfield  {author} {\bibinfo {author} {\bibfnamefont {S.}~\bibnamefont
  {Bose}}\ and\ \bibinfo {author} {\bibfnamefont {V.}~\bibnamefont {Vedral}},\
  }\href {\doibase 10.1103/PhysRevA.61.040101} {\bibfield  {journal} {\bibinfo
  {journal} {Phys. Rev. A}\ }\textbf {\bibinfo {volume} {61}},\ \bibinfo
  {pages} {040101} (\bibinfo {year} {2000})}\BibitemShut {NoStop}%
\bibitem [{\citenamefont {Munro}\ \emph {et~al.}(2001)\citenamefont {Munro},
  \citenamefont {James}, \citenamefont {White},\ and\ \citenamefont
  {Kwiat}}]{Munro2001}%
  \BibitemOpen
  \bibfield  {author} {\bibinfo {author} {\bibfnamefont {W.}~\bibnamefont
  {Munro}}, \bibinfo {author} {\bibfnamefont {D.}~\bibnamefont {James}},
  \bibinfo {author} {\bibfnamefont {A.}~\bibnamefont {White}}, \ and\ \bibinfo
  {author} {\bibfnamefont {P.}~\bibnamefont {Kwiat}},\ }\href {\doibase
  10.1103/PhysRevA.64.030302} {\bibfield  {journal} {\bibinfo  {journal} {Phys.
  Rev. A}\ }\textbf {\bibinfo {volume} {64}},\ \bibinfo {pages} {030302}
  (\bibinfo {year} {2001})}\BibitemShut {NoStop}%
\bibitem [{\citenamefont {Takesue}\ and\ \citenamefont
  {Shimizu}(2010)}]{Takesue2010}%
  \BibitemOpen
  \bibfield  {author} {\bibinfo {author} {\bibfnamefont {H.}~\bibnamefont
  {Takesue}}\ and\ \bibinfo {author} {\bibfnamefont {K.}~\bibnamefont
  {Shimizu}},\ }\href {\doibase 10.1016/j.optcom.2009.10.008} {\bibfield
  {journal} {\bibinfo  {journal} {Opt. Commun.}\ }\textbf {\bibinfo {volume}
  {283}},\ \bibinfo {pages} {276} (\bibinfo {year} {2010})}\BibitemShut
  {NoStop}%
\end{thebibliography}%

%\end{references}

\printfigures

\end{document}